\documentclass[conference]{IEEEtran}
\usepackage{amssymb}
\usepackage{amsmath}
\usepackage{graphicx}
\usepackage[utf8]{inputenc}
\usepackage[numbers]{natbib}
\usepackage[T1]{fontenc}
\usepackage{etoolbox}
\usepackage{flushend}

\newcommand{\SecAgg}{\textsc{SecAgg}}
\newcommand{\mathsc}[1]{{\normalfont\textsc{#1}}}
\DeclareMathOperator{\WrappedNormal}{\mathsc{WrappedNormal}}
\DeclareMathOperator{\Uniform}{\mathsc{Uniform}}
\DeclareMathOperator{\Prob}{\mathbb{P}}

\newcommand\blfootnote[1]{%
  \begingroup
  \renewcommand\thefootnote{}\footnote{#1}%
  \addtocounter{footnote}{-1}%
  \endgroup
}

\title{Federated Learning with Autotuned Communication-Efficient Secure Aggregation}
\author{\IEEEauthorblockN{Keith Bonawitz\IEEEauthorrefmark{1}, Fariborz Salehi \IEEEauthorrefmark{2}, Jakub Konečný\IEEEauthorrefmark{1}, Brendan McMahan\IEEEauthorrefmark{1} and Marco Gruteser\IEEEauthorrefmark{1}}
\IEEEauthorblockA{\IEEEauthorrefmark{1}Google, \{bonawitz, konkey, mcmahan, gruteser\}@google.com}
\IEEEauthorblockA{\IEEEauthorrefmark{2}California Institute of Technology, fsalehi@caltech.edu}
}

\date{November 2019}

\begin{document}

\maketitle

\begin{abstract}
Federated Learning enables mobile devices to collaboratively learn a shared inference model while keeping all the training data on a user's device, decoupling the ability to do machine learning from the need to store the data in the cloud.  Existing work on federated learning with limited communication demonstrates how random rotation can enable users' model updates to be quantized much more efficiently, reducing the communication cost between users and the server.  Meanwhile, secure aggregation enables the server to learn an aggregate of at least a threshold number of device's model contributions without observing any individual device's contribution in unaggregated form.  In this paper, we highlight some of the challenges of setting the parameters for secure aggregation to achieve communication efficiency, especially in the context of the aggressively quantized inputs enabled by random rotation.  We then develop a recipe for auto-tuning communication-efficient secure aggregation, based on specific properties of random rotation and secure aggregation -- namely, the predictable distribution of vector entries post-rotation and the modular wrapping inherent in secure aggregation.  We present both theoretical results and initial experiments.\blfootnote{\IEEEauthorrefmark{2}Work performed while interning at Google.}
\end{abstract}

\section{Introduction}
It is increasingly the case that systems and applications are depending on machine learning models, often deep neural networks, in order to power the features their users require.  Training these machine learning models requires access to data.  In many cases, this training data arises naturally in a distributed fashion, such as on the millions of smartphones with which users interact daily.  In many problem domains, the training data may also be privacy sensitive.  For example, a virtual keyboard application on a smartphone typically requires one or more machine learning models to power features such as tap typing, gesture typing, auto-corrections, and so on.  The most applicable training data for such models are the actual interactions of real users with their virtual keyboards as they live their digital lives.  Because of the potential sensitivity of this training data, there is broad desire for solutions which systematically preserve privacy, for example by ensuring that raw training data never needs to leave the users' devices.

\subsection{Federated Learning with Limited Communication}
Federated Learning addresses this need by enabling mobile devices to collaboratively learn a shared inference model while keeping all the training data on device, decoupling the ability to do machine learning from the need to store the data in the cloud.  In a federated learning system, each user device maintains a local set of private training examples generated by on-device interactions or measurements, while a central server maintains the current version of the model parameters.  For each iteration of model training, the federated learning server selects a cohort of devices from those available for training.  Each device in the cohort downloads a copy of the current model parameters from the server, then uses the devices local training examples to form a model update, i.e. by taking some number of steps of stochastic gradient descent and computing the difference between the  model parameters received from the server and the model parameters after local training.  The server then aggregates the model updates from all devices into average model update, which it then adds to the current model parameters to form a new set of model parameters, ready for then next iteration of training.

In federated learning systems for consumer devices such as smartphones, the devices are interacting with the server over consumer internet connections. While these interactions may be scheduled at times when the consumer internet connections are most reliable and least expensive, e.g. when the device is connected to a broadband internet service while in the user's home, it is still desired to minimize the bandwidth needs as much as possible since these bandwidth needs would add to those of many other device update and maintenance processes. A more communication-efficient secure aggregation technique could also allow training more models or rounds within a given user bandwidth quota.  

Much research has explored how to minimize communication costs during distributed stochastic gradient descent, including in federated learning scenarios. For example, \citep{suresh2017, konevcny2016federated} demonstrate how distributed mean estimation, as used in federated learning to aggregate model updates from user devices, can be achieved with limited communication. They describe a scheme in which the server randomly selects a rotation matrix $R$ for each aggregation round; each user multiplies their update vector by the random rotation matrix before quantizing and submitting for aggregation. The server applies the inverse rotation to the aggregate vector to recover an estimate of the distributed mean. \citet{suresh2017} show that even aggressive quantization benefits greatly from pre-processing with a random rotation: for $n$ users, when the rotated update vector $x^{(u)}$ is quantized to a single bit per dimension, a mean squared error (MSE) of $\Theta\left(\frac{\log d}{n}\cdot\frac{1}{n}\sum_{u=1}^n ||x^{(u)}||^2_2 \right)$ is achieved, compared to $\Theta\left(\frac{d}{n}\cdot\frac{1}{n}\sum_{u=1}^n ||x^{(u)}||^2_2 \right)$ when the same quantization is used without random rotation.  Furthermore, the same MSE is achieved when the random rotation is replaced with a structured random orthogonal matrix $R=HD$, where $H$ is a Walsh-Hadamard matrix~\cite{horadam2012hadamard} and $D$ is a random diagonal matrix with \textit{i.i.d.} Rademacher entries ($\pm 1$ with equal probability), while achieving $O(d \log d)$ computation in $O(1)$ additional space and with $O(1)$ additional communication (for a seed to a PRNG that generates the $D$ matrix). We note that logarithmic dependence on $d$ in the above MSE bound can be replaced with a constant~\cite{caldas2018expanding}, by appropriate use of Kashin's representation~\citep{lyubarskii2010uncertainty}. However, the technique would not be compatible with the statistical analysis that follows.

\subsection{Secure Aggregation}
In order to further preserve users' privacy, federated learning systems can use techniques from trusted computing or secure multiparty computation to ensure that the server only gets to see the aggregate of user cohorts' model updates and learns nothing further about the individual users' model updates. 

\citeauthor{bonawitz2017practical} demonstrate \SecAgg, a practical protocol for secure aggregation in the federated learning setting, achieving $<2\times$ communication expansion while tolerating up to $\frac{1}{3}$ user devices dropping out midway through the protocol and while maintaining security against an adversary with malicious control of up to $\frac{1}{3}$ of the user devices and full visibility of everything happening on the server~\citep{bonawitz2017practical}. The key idea in \SecAgg~is to have each pair of users agree on randomly sampled $0$-sum pairs of mask vectors of the same lengths as the model updates. Before submitting their model update to the server, each user adds their half of each mask-pair that they share with another user; by working in the space of integers $\bmod~k$ and sampling masks uniformly over $[0,k)^d$, \SecAgg~guarantees that each user's masked update is indistinguishable from random value on its own. However, once all the users updates are added together, all the mask-pairs cancel out and the desired value (the sum of users inputs $\bmod~k$) is recovered exactly.  To achieve robustness while maintaining security, \SecAgg~uses $k$-of-$n$ threshold secret sharing to support recovering the pair-wise masks of a limited number of dropped-out users.

Note that model updates are generally real-valued vectors in federated learning, but \SecAgg~(and similar cryptographic protocols) require input vector elements to be integers $\bmod~k$.  In practice, this is typically solved by choosing a fixed range of the real numbers, say $[-t, t]$, clipping each user update $x^{(u)}$ onto this range, then uniformly quantizing the remaining values using $\kappa$ bins, each of width $\frac{2t}{\kappa - 1}$, such that a real value of $-t$ maps to a quantized value of 0 and a real value of $+t$ maps to a quantized value of $\kappa-1$.  Note that in \SecAgg, the same modulus $k$ applies both to the users' individual inputs and to the aggregated vector.  As such, choosing the \SecAgg~modulus to be $k = n\kappa$, where $n$ is the number of users, ensures that all possible aggregate vectors will be representable without overflow~\cite{bonawitz2017practical}.

\section{Autotuning Communication-Efficient Secure Aggregation}
In this Section, we explain why a straightforward combination of \SecAgg~and the compression techniques affects the relative efficiency, and propose a concrete approach which yields better results.

\subsection{Challenges}
We first note that the majority of the bandwidth expansion for \SecAgg~comes from the choice of $k=n\kappa$.  For $n=2^{10}$ users and $\kappa=2^{16}$ (i.e. 16 bit fixed point representation), \citet{bonawitz2017practical} reports $1.73\times$ bandwidth expansion over just sending the quantized input vector in the clear. Some of this bandwidth expansion is associated with secret sharing and other cryptographic aspects of the protocol.  However, observe that choosing $k=n\kappa$ with $n=2^{10}$ means that the \SecAgg~modulus is 10 bits wider than $\kappa$; this alone accounts for $\frac{26}{16}=1.625\times$ bandwidth expansion -- the majority of what is reported.  

If we consider combining \SecAgg~with aggressive quantization, e.g. as described in \citep{suresh2017}, the relative expansion cost becomes even more pronounced, as aggressive quantization reduces $\kappa$ but leaves $n$ unchanged.  In the extreme example of single bit quantization, the relative expansion grows to $11\times$ just to ensure the \SecAgg~modulus can accommodate the sum\footnote{Note that the \textit{absolute} communication overhead remains constant; only the \textit{relative} overhead increases}.

We also observe that quantizing to a fixed point representation requires selecting the clipping range $[-t, t]$ \emph{a priori} -- it needs to be the same for each user and thus the server, or an engineer, chooses an appropriate $t$ before the start of a training round. If the clipping range is set smaller than the dynamic range of the users' model updates, then individual model updates may be distorted due to clipping, thereby distorting the computed aggregate as well.  However, as the clipping range increases, one must either (a) increase the number of quantization bits used, hence driving up the communication cost, or (b) incur a higher variance estimate of the aggregate due to coarser effective quantization.

Establishing an explicit clipping range can be challenging for the model engineer.  Many ML engineers have little intuition about the dynamic range of model updates, in part because that dynamic range can depend on a variety of factors including the neural network architecture, activation functions, learning rate, number of passes through the data per model update, and even vary as training progresses. The dynamic range will typically also vary between different model variables/layers.

As such, we desire an automated means by which the clipping range can be selected.

Secure Aggregation can make this more difficult to determine empirically, because the ML engineer is only able to view the aggregate model update across all users in the round, \textit{after} any distortion from clipping has already occurred on the user devices.  While we could gather additional signals from user devices to facilitate setting the clipping range appropriate, we would prefer not to do so.  \SecAgg~is generally used in order to protect the privacy of the users' input signals; any additional signals would also have to have their privacy properties reasoned about.  For example, if \SecAgg~is being used to facilitate differential privacy, then some portion of the privacy budget would need to be allocated to privacy costs associated with any additional signals gathered for clipping range tuning.

\subsection{Autotuning Overview}

Fortunately, we can take advantage of two unusual properties of \SecAgg~and randomized rotation in order to construct a recipe for automated tuning that requires no additional signals from the user devices. First, modular wrapping in \SecAgg allows users to compute values $\bmod~k$ instead of clipping them, which preserves a signal from the tails of the distribution in the sum. Second, the randomized rotation step produces inputs with a normal distribution that changes based on the degree to which values ``wrap around'' in the modular operation. This allows the server to estimate the original distribution and adjust the quantization range to minimize such wrapping. With this precisely tuned quantization, secure aggregation can then operate with significantly smaller fixed point integer representations and achieve improved communication efficiency.

We consider these properties and the automated tuning recipe further in the next subsections.

\subsection{Modular Wrapping in \SecAgg}
\label{sec:modular-wraping}

Recall that \SecAgg~computes sums $\bmod~k$.  Because $\bmod~k$ is an idempotent operation, and because $\bmod$ and summation commute, we find that each user can compute their input $\bmod~k$ \textit{before} submitting it to secure aggregation without affecting the result at all.  That is, if $x^{(u)}$ is the update from user $u$, and $\mathcal U$ denote the set of all the users participating in an execution of the secure aggregation protocol, then $\SecAgg (\{x^{(u)}\}_{u\in \mathcal U}) = \big(\sum_{u\in \mathcal U} x^{(u)}\big) \bmod~k = \big(\sum_{u\in \mathcal U} \left(x^{(u)} \bmod~k\right)\big) \bmod~k = \SecAgg\big( \{x^{(u)} \bmod~k\}_{u\in \mathcal U}\big)$.

This suggests an alternative to the standard approach of clipping to a fixed range, then quantizing the result.  Instead, we'll consider quantizing first (over an unbounded range), then applying the $\bmod~k$ operation \textit{instead of} clipping.  When we clip before quantizing, distortion is introduced whenever an individual user's contribution exceeds the fixed point range allocated to that individual.  In contrast, by quantizing then applying $\bmod~k$, we only introduce distortion if the \emph{true sum over all users' inputs $\sum_u x^{(u)}$} lies outside the fixed point range allocated to the representation of sum.  

\subsection{Randomized Rotation Produces (Almost) Normally Distributed Inputs}
\label{sec:random-rotation-normal}
When a randomized rotation matrix $R$ is applied to a vector $x$, the entries of $y=Rx$ have identical distribution with mean $0$ and variance equal to $||x||^2_2 / d$. It can be shown that as $d$ grows large, the distribution of each of the entries of the vector $y$ will approach a Gaussian distribution, i.e., $N(0, ||x||^2_2 / d)$. That is, for any input vector $x$, if we form a histogram of the entries in $y$, we expect to see a normal distribution\footnote{For large values of $d$, it can also be shown that the a vector $v\in \mathbb R^d$ with entries drawn independently from the Gaussian distribution $N(0,\frac{1}{d})$ will concentrate around the unit sphere.}.

To show this, we first note that a random rotation $R \in \mathbb{R}^{d\times d}$ is a unitary matrix, with the columns forming an orthonormal basis. Because random rotation is simply representing the vectors in a new basis, it follows immediately that the $\ell_2$-norm is preserved, i.e.  ${||Rx||}_2={||x||}_2$. 

Let $\mathcal O_{d}$ denote the set of all unitary rotation matrices over $\mathbb R^d$, i.e., $\mathcal O_{d}=\{ M\in \mathbb R^{d\times d}: MM^T=M^TM=I_d\}$. Let $v \in {\mathbb S}^{d-1}$ be a vector in the unit sphere, and choose a random matrix $R$ uniformly from the set $\mathcal O_{d}$. We then observe that $Rv$ has a uniform distribution on the unit sphere, i.e., $Rv \sim \Uniform({\mathbb S}^{d-1})$. The same argument gives the following,
%Assuming $R\sim \Uniform(\mathcal O_{d})$, such that $\forall v \in {\mathbb S}^{d-1}, Rv \sim \Uniform({\mathbb S}^{d-1})$, then we have:

\begin{equation}
    y\sim \Uniform({||x||}_2\cdot\mathbb S^{d-1})~,
\end{equation}

where $\mathbb S^{d-1}$ denotes the unit sphere in $\mathbb R^d$.  As a direct consequence of the isoperimetric inequality on the unit sphere~\cite{vershynin2018high}, we find that 

\begin{equation}
    \Prob\{|y_i|>\tau\}\leq 2e^{-\frac{d\tau^2}{2{||x||}_2^2}}
\end{equation}

which implies that the entries of $y$ have a sub-Gaussian distribution.

This applies directly to the randomly rotated model updates $x^{(u)}$ from each user.  However, because summation and matrix multiplication commute, the entries of the sum of the users' rotated values will also be (almost) normally distributed, with $\bar{y}_k \sim N\left(0, \frac{||\bar{x}||^2_2}{d}\right)$ where $\bar{x} = \sum_u x^{(u)}$ and $\bar{y} = \sum_u Rx^{(u)} = R\bar{x}$.  

Similarly, when we replace the random rotation by the randomized Hadamard matrix $R=HD$ as in \citep{suresh2017}, the entries are \emph{approximately} normal, and identically but not independently distributed. Nevertheless, for high dimensional values of interest here, the difference is small and we will see later that the theoretical insights presented next carry over to practice.

% While it is harder to prove, it is easy to demonstrate empirically that this property continues to hold true $R$ is replaced with a structured pseudorandom matrix $R=HD$ as in \cite{suresh2017}.

\subsection{Combining Modularity and Random Rotation}
\label{sec:wrapped-normal}
We have established that if we randomly rotate each input vector $x^{(u)}$ before aggregating, then we expect the sum $\bar{y}$ to have  normally distributed entries.  Observe further that if we randomly rotate $x^{(u)}$ and then stochastically quantize (without clipping or modular wrapping), we still expect the entries of the (dequantized) sum to be approximately normally distributed.

\begin{figure}[!t]
\centering
\includegraphics[width=2.5in]{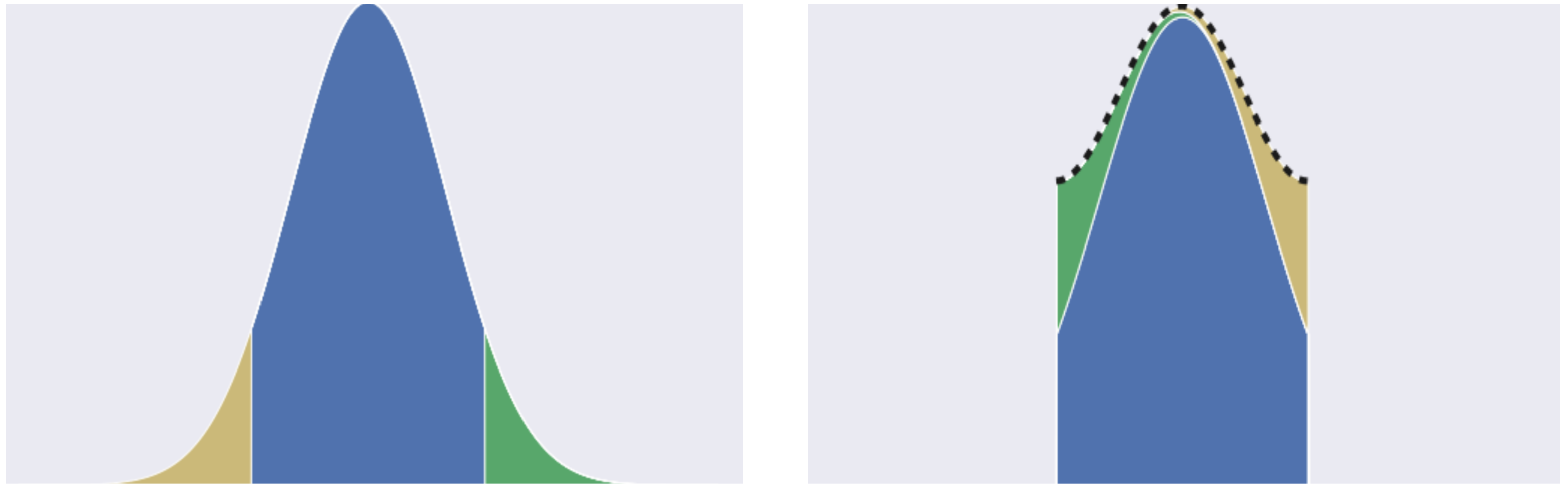}
\caption{The pdf of a normal distribution (left) and a corresponding $\WrappedNormal$ distribution (right).  Note how the green and yellow tails of the normal distribution ``wrap around'' in the $\WrappedNormal$.}
\label{fig:wrappednormal}
\end{figure}

Now consider randomly rotating the $x^{(u)}$, quantizing, then applying the $\bmod~k$ before computing the sum $\bmod~k$, as in the \SecAgg~setting.  After dequantizing the sum, we can no longer expect a normal distribution: the domain of the tails of the normal distribution correspond to values outside the range $[-t, t]$ that maps to the valid quantized values $[0, k-1]$.  Instead, we find that the tails of the normal distribution ``wrap around,'' producing instead a \textit{wrapped normal distribution}~\cite{mardia2009directional}, see Figure~\ref{fig:wrappednormal}, with the probability density function
\begin{multline*}
\WrappedNormal(x; \mu, \sigma)= \\ 
\frac{1}{\sigma\sqrt{2\pi}} \sum_{k=-\infty}^{\infty} \exp\left[ \frac{-(x - \mu + 2 \pi k)^2}{2 \sigma^2}\right]
\end{multline*}
where $\mu$ and $\sigma$ are the mean and variance of the unwrapped normal distribution, and the domain is assumed to be $[-\pi, \pi]$.  The range $[-t, t]$ can be obtained by scaling the distribution appropriately.

\subsection{A Recipe for Autotuning Communication-Efficient \SecAgg}
\label{sec:recipe}

Taken together, sections~\ref{sec:modular-wraping} through \ref{sec:wrapped-normal} suggest a concrete recipe for performing communication-efficient secure aggregation, while autotuning the quantization strategy, and without requiring any additional signals to be communicate to the server. Given $\alpha$, the probability that an individual entry of the rotated sum $\bar{z}$ can be distorted, proceed as follows:

\begin{enumerate}
    \item The server selects a structured pseudorandom rotation matrix $R=HD$ and communicates it to each user.
    \item Each user randomly rotates their input $z^{(u)}=Rx^{(u)}$
    \item Each user quantizes $z^{(u)}$ (over an unbounded range) with the current quantization bin size $b$, then applies the $\bmod~k$ operation to produce the user's \SecAgg~input $y^{(u)}$
    \item The \SecAgg~protocol is run to produce $\bar{y} = \sum_u y^{(u)} \bmod~k$.
    \item The server computes a histogram of the dequantized entries of $\bar{y}$ and fits a $\WrappedNormal(0, \sigma, t)$ distribution of unknown variance $\sigma^2$ to the result.  From this, the server infers that $\bar{z} = \sum_u z^{(u)}$, the sum of the users' contributions as if we hadn't used quantization or modular wrapping in our aggregation, is distributed as $\bar{z}_i \sim N(0, \sigma)$.
    \item The server uses the inverse cdf for $N(0, \sigma)$ to set $t$ such that $\Prob(\bar{z}_k \not\in [-t, t]) \leq \alpha$ for some constant $\alpha$. Recall that, $\alpha$ is the probability that an individual entry of the rotated sum $\bar{z}$ is distorted due to modular wrapping, $(1-\alpha)^d$ is the probability that no entry in $\bar{z}$ is distorted, and $\alpha d$ is the expected number of distorted entries.
    \item The server compute the new quantization bin size $b^{*} = \frac{2t}{k-1}$, such that the range $[-t, t]$ maps to the full set of quantized output values $[0, k-1]$.
    \item The next iteration of federated learning repeats this recipe using the new bin size $b^{*}$.
\end{enumerate}

\subsection{Fitting a Wrapped Normal Distribution}
In order to implement the recipe above, we still require a practical way to fit the wrapped normal distribution to the observed dequantized entries of $\bar{y}$.  

Following~\cite{mardia2009directional}, we observe that a (biased) estimator of $\sigma^2$ for $\bar{y}\sim\WrappedNormal(0, \sigma)$ can be formed by computing
$$
\bar{R}^2 = \left(\frac{1}{d}\sum_{i=1}^{d} \cos \bar{y}_i\right)^2 + \left(\frac{1}{d}\sum_{i=1}^{d} \sin \bar{y}_i\right)^2~,
$$
$$
R_e^2 = \frac{d}{d-1}\left(\bar{R}^2 - \frac{1}{d}\right)~,
$$
$$
\hat{\sigma}^2 = \ln\left(\frac{1}{R_e^2}\right)~.
$$

\section{Experiments}

Following the recipe outlined in section~\ref{sec:recipe} and using the Federated Averaging algorithm for federated learning~\cite{mcmahan2017communication}, we conducted experiments on the CIFAR-10 dataset~\cite{Krizhevsky09learningmultiple}.  CIFAR-10 consists of 50000 training images + 10000 test images, each 32x32 color pixels, balanced across 10 classes (e.g. airplane, automobile, etc.).  We used a version of the all-convolutional neural architecture in~\cite{springenberg2015striving}, the same as used previously for such experiment~\cite{konevcny2016federated}. We simulated 100 devices for federated learning with a balanced \textit{i.i.d.} partition of the training data across devices.  In each round of federated learning, we selected $10\%$ of the devices to participate.  To compute a device's model update, we used a batch size of 10 training examples and made a single pass through the device's data with a learning rate of 0.1.  We fixed the \SecAgg~modulus for the entries of the sum to $k=2^8$, i.e. 8 bit fixed point quantization.

\begin{figure}[!t]
\centering
\includegraphics[width=2.5in]{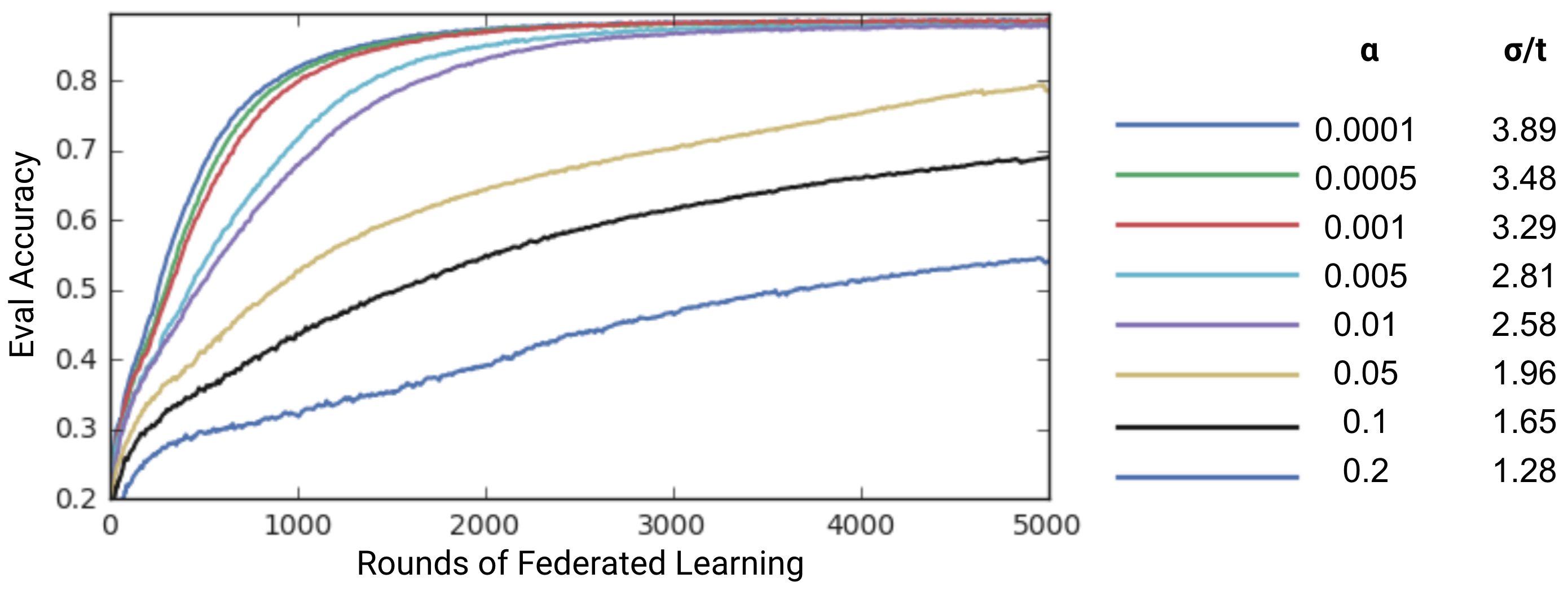}
\caption{Evaluation accuracy for CIFAR-10 experiments, considering various values of $\alpha$.  The legend also lists the equivalent $\frac{\sigma}{t}$ ratio for each $\alpha$.}
\label{fig:cifar10}
\end{figure}

\begin{figure}[!t]
\centering
\includegraphics[width=2.5in]{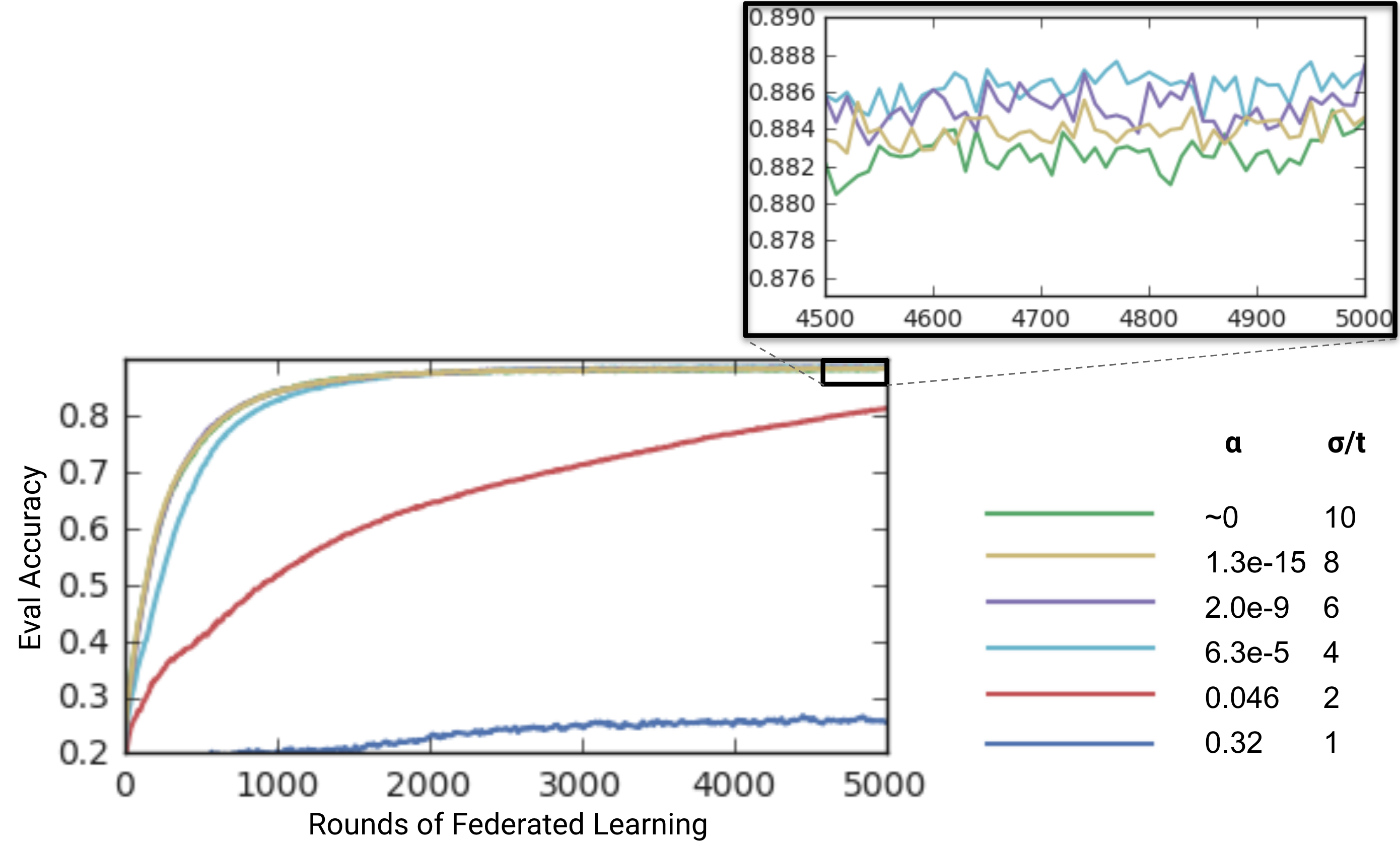}
\caption{Evaluation accuracy for CIFAR-10 experiments.  In
this range of $\alpha$ values, one can see that being too
conservative in the $\alpha$ setting (i.e. driving the probability of modular wrapping towards 0) causes a loss in accuracy, due to overly coarse quantization bins.}
\label{fig:cifar10_zoom}
\end{figure}

Figures~\ref{fig:cifar10} and~\ref{fig:cifar10_zoom} show the results of the experiments, using various values of $\alpha$ for the autotuning recipe. Note that in Figure~\ref{fig:cifar10_zoom}, one can see that being too
conservative in the $\alpha$ setting (i.e. driving the probability of modular wrapping towards 0) causes a loss in accuracy, due to overly coarse quantization bins.

\section{Discussion}

In this paper, we derived and tested a recipe for federated learning with autotuned communication-efficient secure aggregation, with initial results on CIFAR-10 showing promising potential.  

In future experiments, we hope to explore the behavior of the autotuning system on more complex neural networks and further explore the trade off between the number of bits in the \SecAgg~modulus, the setting of $\alpha$, and the achievable machine learning accuracy.  In addition, while we believe it to be easier to reason about $\alpha$ than to accurately guess the dynamic range of the model updates, we also hope to develop a better theoretical understanding of the impact of $\alpha$ on the convergence of the training algorithm.

\bibliographystyle{IEEEtranN}
\bibliography{IEEEabrv,asilomar}

\end{document}